\documentclass[runningheads]{llncs}

\usepackage{amsmath}
\usepackage{graphicx}
\usepackage{hyperref}

\usepackage{booktabs}
\usepackage{framed,multirow}

\usepackage{wrapfig}
\usepackage{ulem}
\usepackage{threeparttable}
\usepackage{array}
\usepackage{bm}
\usepackage{blindtext}
\usepackage{bbding}

\begin{document}

\title{Recurrent Aggregation Learning for Multi-View Echocardiographic Sequences Segmentation}
\titlerunning{Recurrent Aggregation Learning for Multi-View Sequences Segmentation}

\author{
    Ming Li \inst{1,2} \and
    Weiwei Zhang \inst{1} \and
    Guang Yang \inst{3,4} \and
    Chengjia Wang \inst{5} \and
    Heye Zhang \inst{6 (}\Envelope\inst{) } \and
    Huafeng Liu \inst{7} \and
    Wei Zheng \inst{1 (}\Envelope\inst{) } \and
    Shuo Li \inst{8}
    }
\authorrunning{M. Li et al.}

\institute{
    Shenzhen Institutes of Advanced Technology, Chinese Academy of Sciences, Shenzhen, China \\ \email{zhengwei@siat.ac.cn} \and
    Shenzhen College of Advanced Technology, University of Chinese Academy of Sciences, Shenzhen, China \and
    Cardiovascular Research Centre, Royal Brompton Hospital, London SW3 6NP, UK \and 
    National Heart \& Lung Institute, Imperial College London, London SW7 2AZ, UK \and 
    BHF Centre for Cardiovascular Science, University of Edinburgh, Edinburgh, UK \and 
    School of Biomedical Engineering, Sun Yat-Sen University, Shenzhen, China \email{zhangheye@mail.sysu.edu.cn} \and
    Zhejiang University, Hangzhou, China \and 
    Western university, London ON, Canada
    }

\maketitle 

\begin{abstract} 
Multi-view echocardiographic sequences segmentation is crucial for clinical diagnosis.
However, this task is challenging due to limited labeled data, huge noise, and large gaps across views. 
Here we propose a recurrent aggregation learning method to tackle this challenging task.
By pyramid ConvBlocks, multi-level and multi-scale features are extracted efficiently. 
Hierarchical ConvLSTMs next fuse these features and capture spatial-temporal information in multi-level and multi-scale space. 
We further introduce a double-branch aggregation mechanism for segmentation and classification which are mutually promoted by deep aggregation of multi-level and multi-scale features. 
The segmentation branch provides information to guide the classification while the classification branch affords multi-view regularization to refine segmentations and further lessen gaps across views. 
Our method is built as an end-to-end framework for segmentation and classification.
Adequate experiments on our multi-view dataset (9000 labeled images) and the CAMUS dataset (1800 labeled images) corroborate that our method achieves not only superior segmentation and classification accuracy but also prominent temporal stability. 
\end{abstract}

\section{Introduction}
\begin{figure}[t]
    \setlength{\abovecaptionskip}{5pt}
    \centering
    \includegraphics[width=0.67\textwidth]{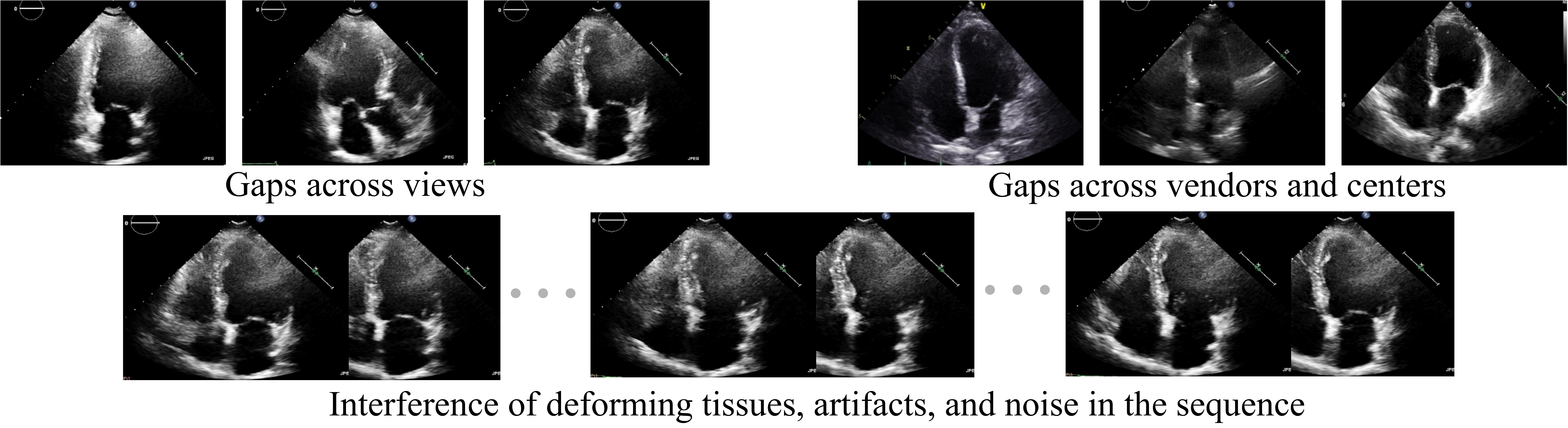}
    \caption{Top left: multi-view samples (A2C, A3C, and A4C). Top right: A4C samples across vendors and centers. Bottom row: echocardiographic sequence.}
    \label{challenge}
\end{figure}

Multi-view echocardiographic sequences delineation provides important insight for clinical diagnosis. 
The knowledge pattern of cardiac structures and textures associated with deforming tissues can be observed in echocardiographic sequence while in single frames the information is always missing and incomplete \cite{huang2014contour}. 
Echocardiographic sequence also permits the assessment of wall motion and identification of end-diastolic (ED) and end-systolic (ES) phases. 
Cardiologists usually check multi-view echocardiographic sequences in clinical decision-making \cite{madani2018fast}. 
The apical-2-chamber view (A2C), A3C, and A4C are the most commonly used views for the left ventricle (LV) functional assessment.
Most clinical indexes of the LV (e.g., area, volume, and ejection fraction) are basically measured in these standard apical views. 
Segmentation of the LV is generally a prerequisite for such quantitative analysis \cite{lang2015recommendations}.  
In clinical routine, quantitative analysis of the LV still involves careful review and massive manual interpretation by experts, which is a tedious and time-consuming task.
Thus, automatic methods are desired to facilitate this process. 
However, multi-view echocardiographic sequences segmentation remains a challenging task as illustrated in Fig. \ref{challenge}. 
First, the fuzzy border, huge noise, and abounding artifacts of echocardiographic images result in local missing and incomplete of the anatomical structures;
Second, multi-view heterogeneous data varies in the anatomical structure, and image properties differ widely across vendors and centers;
Third, in the sequence, artifacts and noise are much severer, and the motion of mitral valve, trabeculation, and papillary muscles also poses additional interference;
Finally, limited labeled data restricts the performance of supervised learning based methods.

The application scenario of existing methods is always limited and only suitable under a specific situation. 
They mostly focus on specific view \cite{carneiro2012segmentation} or single frames (i.e., without considering the sequence) \cite{chen2016iterative} or one single vendor and center \cite{leclerc2019deep}. 
As for sequence segmentation, existing methods try to leverage temporal information by using a deformable model combined with the optical flow \cite{pedrosa2017fast,zhang2019deep} or fine-tuning pretrained CNN dynamically with first frame's label till the last frame \cite{yu2017segmentation}.
The major downsides of these temporal methods are that they are computational cumbersome and not an end-to-end manner. 
The limited labeled data and specific application scenario confine the performance of existing methods and lead to the suboptimal solution. 
\par
\begin{figure}[t]
    \setlength{\abovecaptionskip}{5pt}
    \centering
    \includegraphics[width=0.63\textwidth]{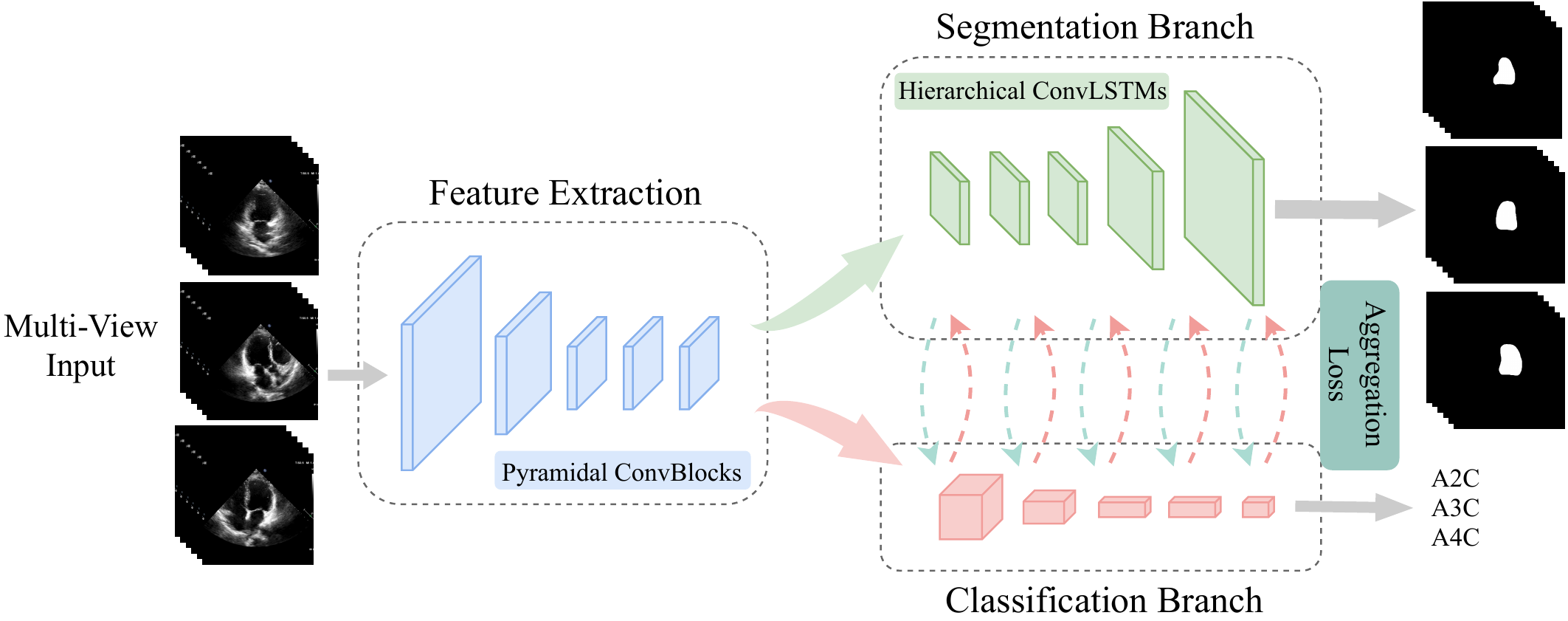}
    \caption{Workflow overview of our method.}
    \label{workflow}
\end{figure}

  
To achieve a unified model for multi-view echocardiographic sequences segmentation, we propose a recurrent aggregation learning method (RAL). The workflow is depicted in Fig. \ref{workflow}. 
Pyramid ConvBlocks joint hierarchical ConvLSTMs are utilized to capture multi-level and multi-scale spatial-temporal information, enabling RAL the ability to harness the knowledge across heterogeneous data (multi-view, multi-center, and multi-vendor). 
We further introduce a double-branch aggregation mechanism for segmentation and classification to lessen gaps across multi-view data. 
Different from existing methods, RAL fully exploits the long term spatial-temporal information in an end-to-end manner and does not depend on any deformable model or optical flow or pretrained segmentation models. 
RAL can accommodate heterogeneous data, not only generate accuracy segmentation results but also achieve the classification of different views at the same time and gain prominent temporal stability.

\section{Method}
\begin{figure}[t]
    \setlength{\abovecaptionskip}{5pt}
    \centering
    \scalebox{0.93}{
    \begin{minipage}[t]{0.12\textwidth}
    \centering
    \includegraphics[width=13mm]{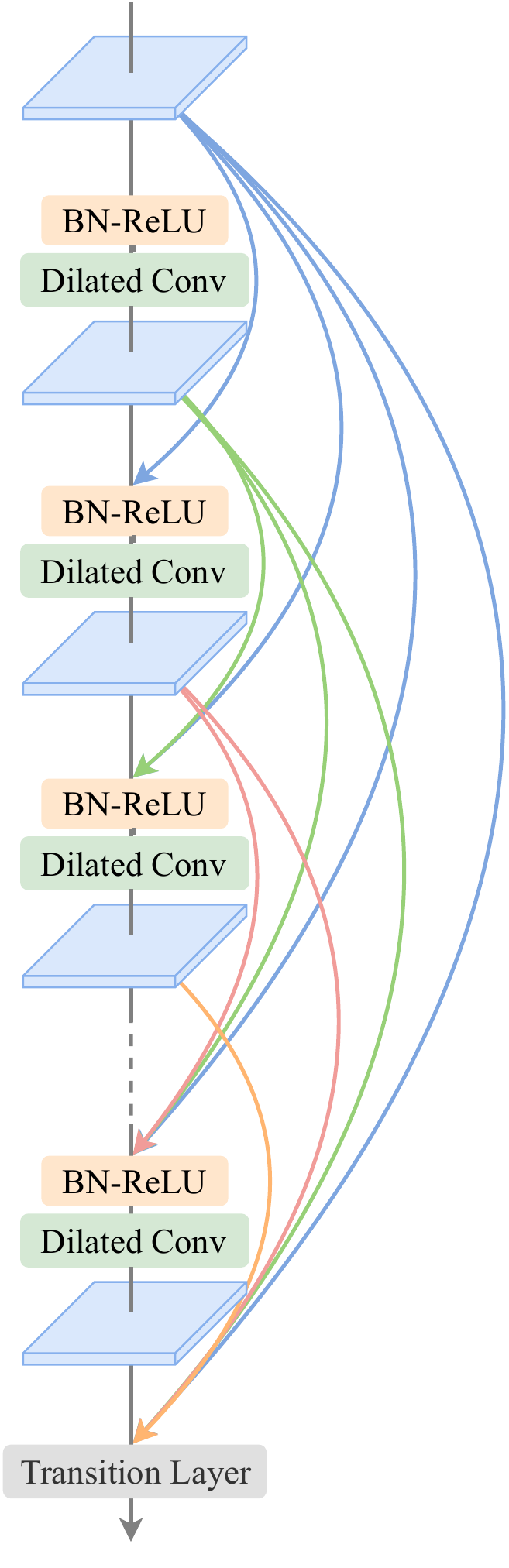}
    \end{minipage}
    \begin{minipage}[t]{0.87\textwidth}
    \centering
    \includegraphics[width=100mm]{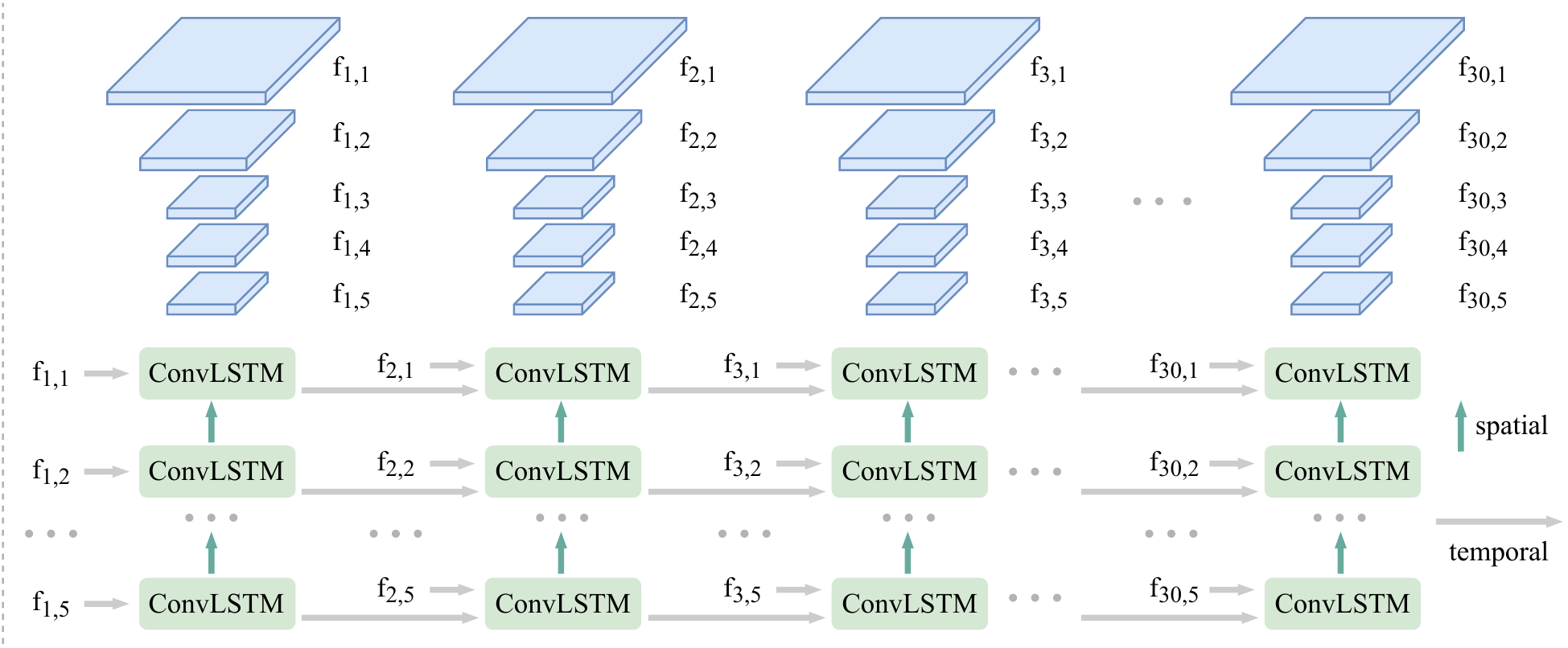}
    \end{minipage}}
    \caption{Left: Dilated dense convolution block. Right: Hierarchical ConvLSTMs for spatial-temporal modeling. }
    \label{detail} 
\end{figure}

RAL is built as an end-to-end framework and comprised of three key components:
the feature extraction module, the segmentation branch, and the classification branch (as depicted in Fig. \ref{workflow}). 
The feature extraction module consists of pyramid dilated dense convolution blocks (ConvBlocks).
The segmentation branch contains hierarchical recurrent architecture of multiple ConvLSTMs \cite{xingjian2015convolutional}.
While the classification branch involves a series of aggregation downsample and fully connected layers.
\newline
\textbf{Multi-Level and Multi-Scale Features Extraction.} 
We design pyramid ConvBlocks architecture in the feature extraction module, which includes 5 ConvBlocks to extract multi-level and multi-scale features.
Multi-level information provides the global geometric characteristic of the LV, while multi-scale information can help to strengthen thin and small regions, further refine the boundaries of the LV. 
They contribute to lessening the gap across views, vendors, and centers, increasing robustness to images conditions and the anatomical structure variations.
One ConvBlock contains $L$ densely connected dilated convolution layers as shown in Fig. \ref{detail}, which can expand the receptive field and meanwhile preserve the resolution of feature maps.
While the transition layer changes channels and resolution of feature maps by convolution and pooling.
The feedforward information propagation from preceding $l$ layers to $(l+1)^{th}$ layer can be formulated as 
\begin{flalign}
    y_{l} = D(C(y_{1},y_{2},...,y_{l-1})) 
\end{flalign}
where $y_{l}$ are the output of the $l^{th}$ layer, $C(\cdot)$ refers to the concatenation of previous layers' outputs. $D(\cdot)$ is a composite function of three connected operations: batch normalization (BN), rectified linear unit (ReLU), and dilated convolution.
Five ConvBlocks generate multi-level and multi-scale features $f_{t}=\{f_{t,1},f_{t,2},f_{t,3},f_{t,4},f_{t,5}\}$ for frame $t$ in the sequence. 
\par
Pyramid ConvBlocks endow RAL with the superior feature extraction ability and the LV region detection capacity in multi-level and multi-scale space, further contribute to capturing the global geometric characteristic of the LV and then establishing uniform semantic features.
Thus RAL can detect and extract the LV accurately and robustly from not only ED and ES frames but also other frames in the sequence where the boundary is not clear (disturbed by noise and other tissues, see sequence samples in Fig. \ref{challenge}).
\newline
\textbf{Recurrent Features Fusion for Spatial-Temporal Modeling.}
\label{sequence}
For sequence segmentation, capturing the LV characteristic over time is essential for temporal stability.
Recent studies based on LSTM have shown great ability to learn sequential information.
Inspired by \cite{chen2018multiview,yang2018multiview}, we conduct hierarchical ConvLSTMs to exploit long term spatial-temporal modeling as depicted in Fig. \ref{detail}.
We add recurrence in the temporal domain to generate prediction $S_{t}$ for frame $t$ in the sequence, which carries forward the LV information from previous frames to following frames and allows the matching between consecutive frames naturally. 
Additionally, we also add recurrence in the spatial domain for multi-level and multi-scale features fusion, which helps to integrate multi-level and multi-scale features efficiently. 
\par
The output $y_{t,k}$ of the $k^{th}$ ConvLSTM at frame $t$ depends on the following variables: 
(1) $k^{th}$ level and scale feature $f_{t,k}$ from the feature extraction module;
(2) the output $y_{t,k-1}$ of preceding $(k-1)^{th}$ ConvLSTM at the same frame $t$;
(3) the output $y_{t-1,k}$ from the $k^{th}$ ConvLSTM of previous frame $t-1$;
(4) the hidden state representation $h_{t,k-1}$ from preceding $(k-1)^{th}$ ConvLSTM at the same frame $t$, which is the spatial hidden state;
(5) the hidden state representation $h_{t-1,k}$ from the $k^{th}$ ConvLSTM of previous frame $t-1$, which is the temporal hidden state. The information flow can be formulated as 
\begin{gather}
x_{input} = [f_{t,k}\ |\ B(y_{t,k-1})\ |\ y_{t-1,k}] \\
h_{state} = [h_{t,k-1}\ |\ h_{t-1,k}] \\
y_{t,k} = ConvLSTM_{k}(x_{input}, h_{state})
\end{gather}       
where $B(\cdot)$ is the bilinear upsampling operator.
At each time step, every ConvLSTM accepts hidden states and encoded spatial-temporal features from previous ConvLSTMs and frame, the corresponding extracted feature from the feature extraction module, it then outputs encoded spatial-temporal features to next ConvLSTM and frame.
Finally, predictions $S_{t}$ are generated by the last ConvLSTM at every frame.
\newline
\textbf{Double-Branch Aggregation Learning.}
To further lessen the gaps across multi-view and refine multi-view segmentation results, we introduce a double-branch aggregation mechanism for simultaneous segmentation and classification of multi-view echocardiographic sequences as depicted in Fig. \ref{workflow}.   
Feature from the last ConvBlock is sent to the classification branch.
Next, it goes through successive convolution and pooling operators to deeply aggregate with multi-level and multi-scale spatial-temporal features from the segmentation branch.
Finally, the classification result is produced by fully connected layers.
\par
The segmentation branch generates multi-view segmentations while the classification branch discriminates the specific view. 
They are mutually promoted by deep aggregation of multi-level and multi-scale spatial-temporal features.
The segmentation branch provides multi-level and multi-scale spatial-temporal information to guide the classification while the classification branch affords multi-view discriminative regularization to refine the segmentation results and further lessen the gaps across views.
This double-branch aggregation mechanism endows RAL outstanding ability to adapt complex variations of anatomical structure.
\par
Additionally, we propose an aggregation loss to dynamically facilitate the communication between the segmentation branch and the classification branch as illustrated in Fig. \ref{workflow}.
The aggregation loss comprises the segmentation loss and classification loss.
The segmentation loss is a combination of binary cross-entropy loss and dice loss. 
While the classification loss is categorical cross-entropy loss. 
Thus the aggregation loss function can be formulated as 
\begin{gather}
    L_{segmentation} = -[G\cdot log(P) + (1-G)\cdot log(1-P)] + \frac{2\cdot G \cdot P}{G + P} \\
    L_{classification} = - \sum_{i=1}^{3}g_{i}\cdot log(p_{i}) \\
    L_{aggregation} = \lambda_{s} \cdot L_{segmentation} + \lambda_{c} \cdot L_{classification} 
\end{gather}  
where $G$ and $P$ denote ground truth and prediction of segmentation respectively, $g$ and $p$ refer to ground truth and prediction of classification separately, $i$ indicates the type of view. 
Besides, $\lambda_{s}$ and $\lambda_{c}$ are the corresponding balance coefficients, both are chosen empirically during the training process. 
\begin{table}[t]
    \setlength{\belowcaptionskip}{5pt}
    \centering
    \caption{Specifications of our dataset (left) and the CAMUS dataset (right). }
    \begin{minipage}[t]{0.63\textwidth}
    \centering
    \scalebox{0.73}{
    \begin{tabular}[c]{cccccccc}
        \toprule
        Vendor Machines & Patients & Sequences & Images & ~ & A2C & A3C & A4C \\
        \midrule
        Philips EPIQ 7C & 60 & 180 & 5400 & \textbf{Sequences} & 100 & 100 & 100\\
        GE VIVID E9 & 20 & 60 & 1800 & Training & \multicolumn{3}{c}{240} \\
        Philips IE33 & 20 & 60 & 1800 & Testing & \multicolumn{3}{c}{60} \\
        \midrule
        Total & 100 & 300 & \textbf{9000} & Total & \multicolumn{3}{c}{300} \\
        \bottomrule
    \end{tabular}}    
    \end{minipage}  
    \begin{minipage}[t]{0.36\textwidth}
    \centering 
    \scalebox{0.73}{
    \begin{tabular}[c]{cccccc}
        \toprule
        CAMUS & ~ & A2C & A4C & Vendor Machine \\
        \midrule
        \textbf{Images} & ~ & 900 & 900 & \multirow{3}*{GE VIVID E95} \\
        Training & ~ & \multicolumn{2}{c}{1600} & \\
        Testing & ~ & \multicolumn{2}{c}{200} & \\
        \midrule
        Total & ~ & \multicolumn{2}{c}{\textbf{1800}} & \\
        \bottomrule
    \end{tabular}}   
    \end{minipage}  
    \label{data}
\end{table}
\section{Experiments}
\begin{figure}[t]
    \setlength{\abovecaptionskip}{5pt}
    \centering
    \includegraphics[width=0.8\textwidth]{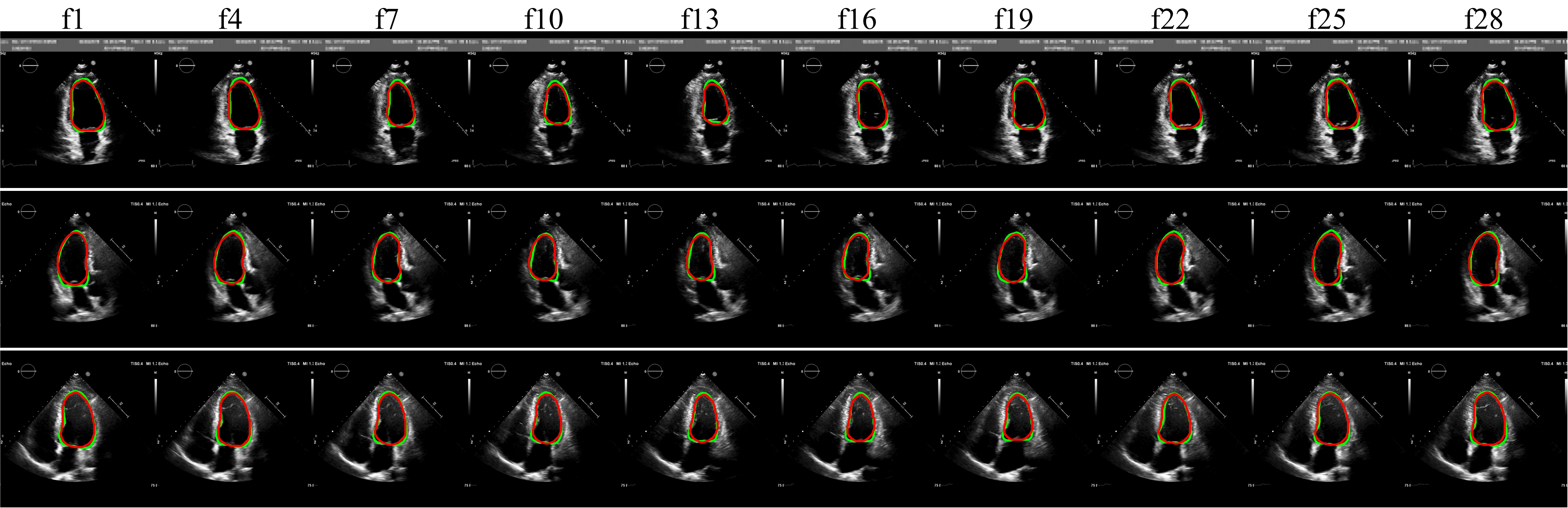}
    \caption{The LV contours of multi-view sequences segmented by our method (red) and experts (green). Ten frames are selected from every sequence to fit the layout view. (top row: A2C; middle row: A3C; bottom row: A4C)}
    \label{segs}
\end{figure}

\textbf{Datasets.}
To validate the efficiency of RAL, we built a large multi-view echocardiographic sequences dataset, which was acquired from three centers' various vendor machines (The Second People's Hospital of Shenzhen, The Third People's Hospital of Shenzhen, and Peking University First Hospital). 
We further evaluate RAL on the public CAMUS dataset \cite{leclerc2019deep}. 
Our dataset contains 300 sequences from 3 views and every sequence includes 30 frames. 
All 9000 frames were labeled by two experts. 
Fig. \ref{segs} presents A2C, A3C, and A4C sequences samples segmented by RAL and experts. 
While the CAMUS dataset only contains manual labels at ED and ES frames, which was acquired from a single vendor and center. 
Table. \ref{data} shows the specifications of two datasets.
\newline
\textbf{Evaluation Metrics.}
Accuracy, Dice, Mean Absolute Distance (MAD) and Hausdorff Distance (HD) are used to measure segmentation results. 
We further evaluate the segmentation performance with the ED, ES volume, and ejection fraction on the CAMUS dataset. 
We utilize the output of RAL to compute clinical indices according to standard guidelines \cite{lang2015recommendations}.
\newline
\textbf{Implementation Details.}
All images are resized to $256\times256$ for computational efficiency.
We employ Adam with the learning rate of 0.001 as the optimizer.
The dilated rates of 5 ConvBlocks are $1, 1, 2, 4, 8$ respectively, and every ConvBlock contains 6 layers.
Besides, a dynamical decay mechanism is utilized to reduce the learning rate by monitoring the change of Dice. 
Ten-fold cross-validation was utilized to provide an unbiased estimation. 
\begin{table}[t]
    \setlength{\belowcaptionskip}{5pt}
    \centering
    \caption{Ablation results of our methods under different configurations.}
    \scalebox{0.83}{
    \begin{tabular}[c]{ccccccc}
        \toprule
        Configurations & Accuracy & Dice & HD(mm) & MAD(mm) & Classification\\
        \midrule 
        full 
        & 0.987$\pm$0.005 & 0.919$\pm$0.040 & 5.87$\pm$3.46 & 2.90$\pm$1.49 & 0.933 \\
        w/o classification 
        & 0.971$\pm$0.008 & 0.910$\pm$0.049 & 5.99$\pm$3.69 & 3.10$\pm$1.66 & - \\
        w/o ConvBlock 
        & 0.963$\pm$0.015 & 0.907$\pm$0.057 & 6.21$\pm$4.95 & 3.27$\pm$1.85 & 0.867 \\
        w/o temporal 
        & 0.955$\pm$0.019 & 0.896$\pm$0.062 & 6.64$\pm$5.04 & 3.51$\pm$1.93 & 0.917 \\
        w/o spatial 
        & 0.968$\pm$0.011 & 0.911$\pm$0.054 & 6.03$\pm$4.16 & 3.08$\pm$1.71 & 0.883 \\
        \bottomrule
    \end{tabular}}
    \label{ablation}      
\end{table}
\newline
\textbf{Ablation Study.}
We evaluate our method under different configurations to corroborate the necessity of every component in RAL. 
The classification branch, ConvBlock, spatial modeling and temporal modeling are removed respectively. 
Table. \ref{ablation} shows the ablation results, we can see that full RAL achieves higher mean values of Accuracy and Dice, lower mean values of HD and MAD, and lower standard deviations of all metrics compared against other configurations. 
RAL also achieves the best classification accuracy (0.933).
Every single component brings important improvement for the LV segmentation, especially when adding recurrence in the temporal domain. 
\newline
\textbf{Comparison Study I: Geometrical.}
We compare RAL with U-Net, ACNN, and U-Net++ on our multi-view sequences dataset.
As shown in Table. \ref{compare}, RAL outperforms other methods on all metrics, achieving the highest mean values of Accuracy (0.987) and Dice (0.919), the lowest mean values of HD (5.87mm) and MAD (2.90mm), and significantly lower standard deviations of all metrics. 
These strongly prove that RAL is able to accomplish the best region coverage, the highest contour accuracy, and the minimum distance error when processing multi-view echocardiographic sequences across multi-vendor and multi-center.
\begin{table}[t]
    \setlength{\belowcaptionskip}{5pt}
    \centering
    \caption{Geometrical comparison results on our multi-view echocardiographic sequences dataset.}
    \scalebox{0.8}{
    \begin{tabular}[c]{cccccc}
        \toprule
        Methods & Accuracy & Dice & HD(mm) & MAD(mm) \\
        \midrule
        RAL & 0.987$\pm$0.005 & 0.919$\pm$0.040 & 5.87$\pm$3.46 & 2.90$\pm$1.49 \\
        U-Net    & 0.942$\pm$0.030 & 0.883$\pm$0.068 & 8.94$\pm$6.87 & 3.72$\pm$1.87 \\
        ACNN     & 0.959$\pm$0.013 & 0.893$\pm$0.061 & 7.70$\pm$6.58 & 3.40$\pm$1.57 \\
        U-Net++  & 0.937$\pm$0.032 & 0.880$\pm$0.072 & 9.01$\pm$7.14 & 3.86$\pm$2.01 \\
        \bottomrule  
    \end{tabular}}     
    \label{compare}  
\end{table}
\begin{figure}[t]
    \setlength{\abovecaptionskip}{2pt}
    \begin{minipage}[t]{0.49\textwidth}
    \centering
    \includegraphics[width=42mm]{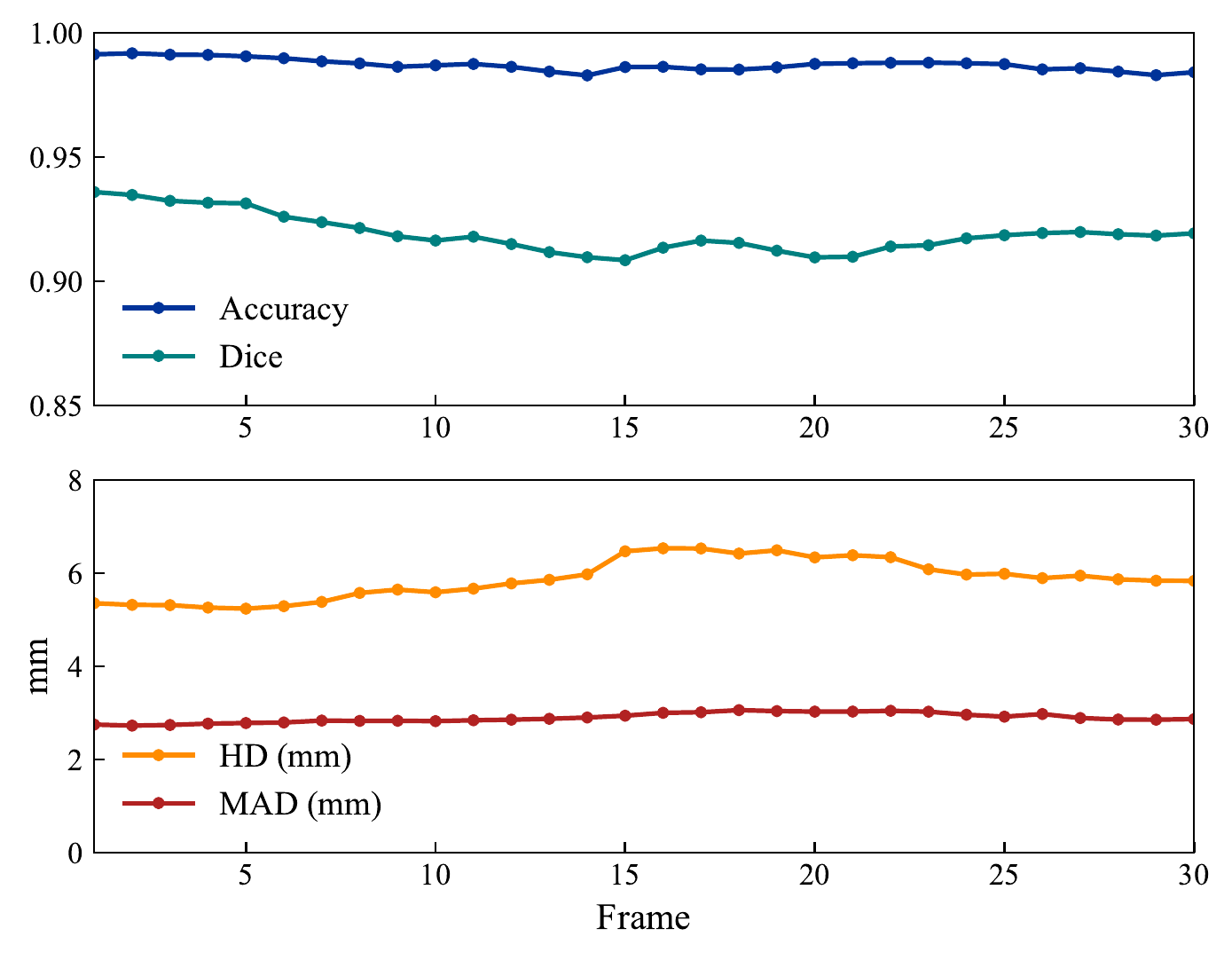}
    \end{minipage}
    \begin{minipage}[t]{0.49\textwidth}
    \centering
    \includegraphics[width=43mm]{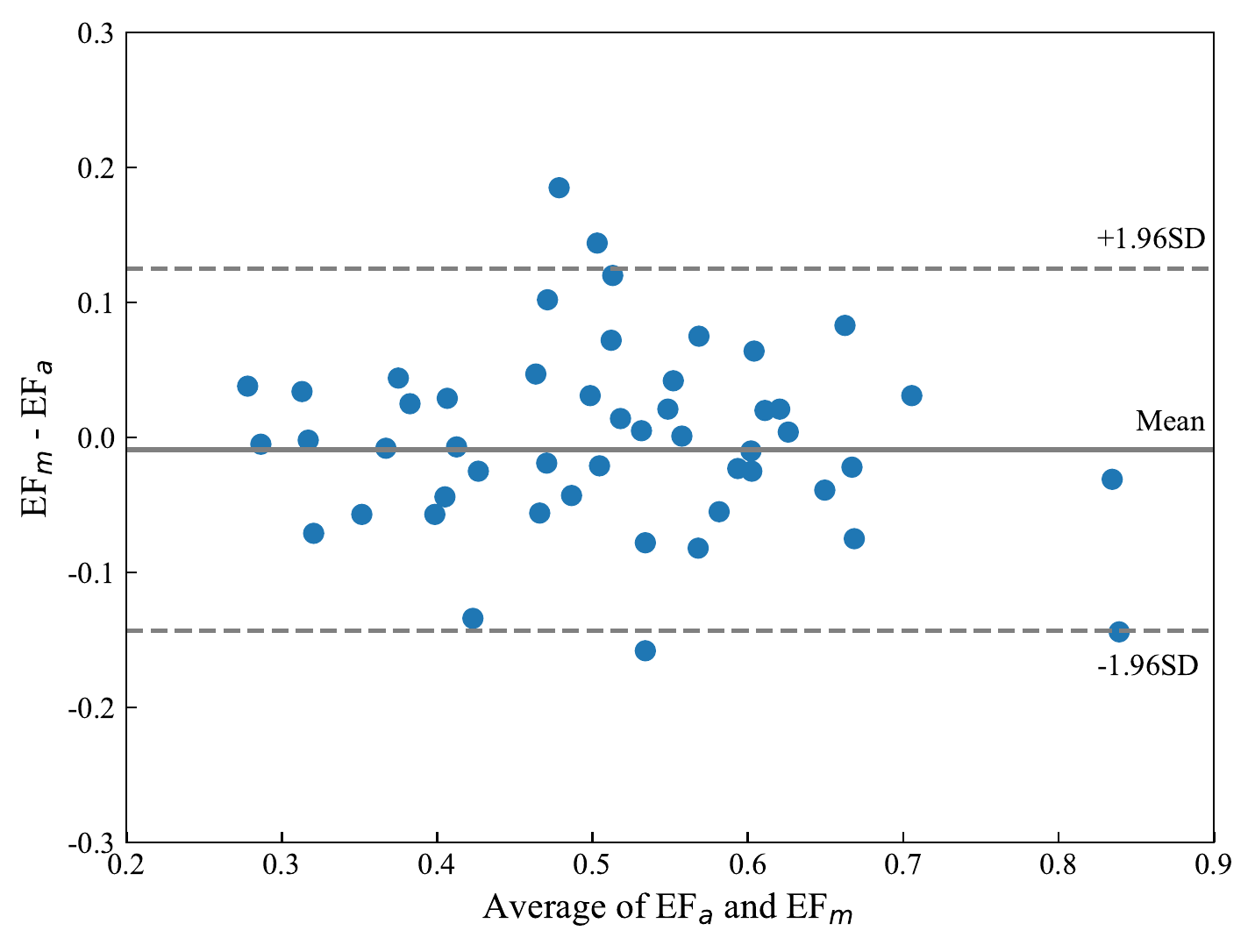}
    \end{minipage}
    \caption{Left: Mean of Accuracy, Dice, HD, and MAD at different frames of the cardiac cycle. Right: Bland–Altman analysis (EF$_a$ and EF$_m$: ejection fraction calculated from automatic segmentations and manual labels) }
    \label{bland_frames}
\end{figure}
\newline
\textbf{Comparison Study II: Clinical.}
We compare RAL with U-Net, ACNN, and U-Net++ on the CAMUS dataset to calculate clinical indices.
As shown in Table. \ref{camus}, 
RAL obtained high correlation scores (0.952 for EDV, 0.960 for ESV, and 0.839 for EF), reasonably small biases and standard deviations, and relatively low mae (8.8ml for EDV, 7.1ml for ESV, and 5.0\% for EF).
Fig. \ref{bland_frames} presents a more intuitional result by Bland–Altman plot.
94\% of the measurements locate in the $\pm$1.96 standard deviation in Bland–Altman plot. 
These results reveal the clinical potential of RAL. 
\newline
\textbf{Temporal Stability.}
We compute the mean of Accuracy, Dice, HD, and MAD at different frames of all echocardiographic sequences and then observe the volatility of each metric to assess the temporal stability.
As shown in Fig. \ref{bland_frames}, RAL achieves stable mean values of all four metrics in the cardiac cycle, only exists moderate fluctuating in the middle of the sequence.
This means spatial-temporal modeling of RAL is efficient. 
RAL achieves not only superior segmentation accuracy but also a good coherence of consecutive frames in the sequence.
\newline
\textbf{Limitation.}
In Fig. \ref{bland_frames}, from ED to ES frames, we observe that Accuracy and Dice decay slightly while HD and MAD increase mildly, and all metrics keep relatively stable in the diastole but show feeblish recoverability.
The sequential process carries errors forward resulting in accumulation of temporal errors in the cardiac cycle. 
Fortunately, the fluctuating rate is moderate and the worst results are still fairly good.
This limitation could be alleviated via Bi-direction LSTM.

\begin{table}[t]
    \setlength{\belowcaptionskip}{5pt}
    \centering
    \caption{Clinical comparison results on CAMUS dataset. (EDV: ED volume; ESV: ES volume; EF: ejection fraction; corr: Pearson correlation; mae: mean absolute error)} 
    \scalebox{0.73}{
    \begin{tabular}[c]{cccccccccccc}
        \toprule
        \multirow{2}*{Methods} & \multicolumn{3}{c}{EDV} & ~ & \multicolumn{3}{c}{ESV} & ~ & \multicolumn{3}{c}{EF} \\
        ~ & corr & bias(ml) & mae(ml) & ~ 
        & corr & bias(ml) & mae(ml) & ~ 
        & corr & bias(\%) & mae(\%) \\ 
        \midrule
        RAL 
        & 0.952 & -7.5$\pm$11.0 & 8.8 & ~
        & 0.960 & -3.8$\pm$9.2 & 7.1 & ~ 
        & 0.839 & -0.9$\pm$6.8 & 5.0 \\
        U-Net    
        & 0.954 & -6.9$\pm$11.8 & 9.8 & ~
        & 0.964 & -3.7$\pm$9.0 & 6.8 & ~ 
        & 0.823 & -1.0$\pm$7.1 & 5.3 \\
        ACNN     
        & 0.945 & -6.7$\pm$12.9 & 10.8 & ~ 
        & 0.947 & -4.0$\pm$10.8 & 8.3 & ~
        & 0.799 & -0.8$\pm$7.5 & 5.7 \\
        U-Net++  
        & 0.946 & -11.4$\pm$12.9 & 13.2 & ~ 
        & 0.952 & -5.7$\pm$10.7 & 8.6 & ~ 
        & 0.789 & -1.8$\pm$7.7 & 5.6 \\
        \bottomrule 
    \end{tabular}}     
    \label{camus}   
\end{table}
\section{Conclusion}
In this paper, we present a recurrent aggregation learning method to exploit long term spatial-temporal information for simultaneous segmentation and classification of multi-view echocardiographic sequences.
Multi-level and multi-scale features are recurrently aggregated on both spatial domain and temporal domain for effective spatial-temporal modeling.
A double-branch aggregation mechanism further brings multi-view discriminative regularization to refine the segmentation results. 
Adequate experiments of geometrical and clinical evaluation demonstrate that RAL achieves not only superior segmentation and classification accuracy, prominent temporal stability, but also high correlations on clinical indices. 
\newline
\\
\textbf{Acknowledgment.}
This work is funded by the Shenzhen Basic Research Program (JCYJ20170818164343304, JCYJ20180507182432303). 

\end{document}